# On Emerging Paradigm of Teaching Measurement Science and Technology in Times of Ubiquitous Use of AI Tools

**Roman Z. Morawski**

**Institute of Radioelectronics & Multimedia Technology, Faculty of Electronics & Information Technology**
**Warsaw University of Technology, Poland**
*roman.morawski@pw.edu.pl*

*Abstract*: The ubiquitous use of the tools of artificial intelligence (AI) in technoscience, in higher education and in other existing and potential fields of measurement application generates new challenges for teaching measurement science and technology (MS&T). The aim of this article is to encourage its readers to modernise their approach to teaching MS&T in a way as to meet these challenges, in particular – to profit from new technological opportunities, and to respond to the needs of our civilisation, identified within a humanistic reflection over its development. First, the state of the art in applications of AI in higher education is briefly characterised. Next, a methodology for using AI tools in MS&T, referring to the author's meta-model of measurement, is outlined. Finally, conclusions concerning an emerging paradigm of teaching MS&T are summarised. The most important of them are as follows. The challenges implied by the ubiquitous use of AI tools cannot be effectively faced without enhancing, in the corresponding curricula, of the contents related to mathematical modelling of material entities, on the one hand, and of the contents related to ethics of research and engineering, on the other. Knowledge and skills related to the art of mathematical modelling are indispensable for extensively profiting from the convergence of various technologies around IT tools, including AI tools. The knowledge and skills related to ethics of research and engineering are indispensable for developing safe applications of measurement in such domains as autonomous vehicles, social robots, biomedical engineering or automated manufacturing.

## 1. Introduction

Explaining the key concepts, used in the title of an article, is always a good practice, but it is of particular importance if those concepts may be interpreted in various ways. This is the case of the concepts of *measurement science and technology* (MS&T) and *artificial intelligence* (AI). The first of them denotes the body of knowledge concerning all aspects of measurement: its philosophical, logical and theoretical foundations, as well as methodologies for its practical implementation, including the design and exploitation of measurement instrumentation.

The lexical definition of the concept of *artificial intelligence* – satisfactory from logical point of view – is impossible since no undisputable definition of the concept of natural human intelligence is available (a collection of 70 attempted definitions may be found in [1]). Consequently, for the time being, an ostensive definition seems to be a viable alternative. The one proposed by Chris Ategeka in his 2022 book [2] can be very helpful in understanding the essence of AI. It consists in distinguishing four types of AI systems, *viz.*:

(1) the systems designed for solving specific tasks (*e.g.* playing chess) which, when solving a new task, do not use the experience acquired during solving similar tasks in the past;
(2) the systems designed for solving specific tasks (*e.g.* driving cars) which, when solving a new task, use the experience acquired during solving similar tasks in the past;
(3) the systems equipped with the so-called social intelligence that are able to read the intentions and predict the behaviour of people (*e.g.* the behaviour of elderly persons they "take care of");
(4) the systems equipped with self-awareness that are able to understand their internal processes imitating human thinking and emotions, as well as intentions and consequences of their actions.

According to the *EU Artificial Intelligence Act*, promulgated in 2024, an AI system is "*a machine-based system designed to operate with varying levels of autonomy, that may exhibit*

*adaptiveness after deployment and that, for explicit or implicit objectives, infers, from the input it receives, how to generate outputs such as predictions, content, recommendations, or decisions that can influence physical or virtual environments*" [3]. Many more ostensive definitions of AI, based on the specification of the subject of this domain of information technology (IT), may be found in the tables of contents of encyclopaedias, handbooks and monographs devoted to AI.

The intuitions concerning the potential behind AI tools when applied in MS&T were verbalised already in the 1980s. They inspired the organisers of the IMEKO TC7 Symposium on Intelligent Measurement held in Jena (GDR) in 1986, whose outcomes were summarised in an 1987 article published in *Measurement* [4]. Since then, review articles on actual and potential applications of AI in MS&T have become increasingly common. Two of them, [5] and [6], are worth being mentioned here since they played an important role in vulgarisation of this topic – both in the community of MS&T researchers and beyond. After 2015, such articles started to appear more often; here are four recent examples: [7–10]. Of course, much more numerous than general articles on the use of AI in MS&T are works devoted to specific applications, especially those related to biomedical engineering, technical diagnostics and industrial monitoring. Here are some recent examples representative of biomedical engineering: [11–22]; and of the other two categories: [23–26]. Of course, numerous examples may be found in other domains of MS&T; here are five representative papers published in 2023: [27–31]. A key aspect of the currently developed applications of AI tools in MS&T is their responsible use embodied in the concept of *trustworthy AI – cf.*, for example, [32]. An AI tool is considered trustworthy if it is able (1) to carry out operations free of bias and discrimination, and to ensure equitable treatment for all users (*fairness*), (2) to provide credible explanations of decisions made (*transparency*), (3) to protect user data (*privacy*), (4) to ensure secure operations (*security*), (5) to function correctly and consistently, also in complex or unexpected situations (*reliability*), and (6) to allow human users to supervise its operation and take responsibility for its actions (*accountability*). A more elaborated definition of trustworthy AI, together with the guidelines for using this definition in practice, may be found in the EC document titled *Assessment List for Trustworthy Artificial Intelligence* [33].

A brief overview of applications of AI tools in MS&T, provided in the previous paragraph, is intended only to demonstrate their direct importance for this branch of technoscience, while the principal aim of this article is rather to encourage its readers to modernise their approach to teaching MS&T in a way as to meet challenges resulting from the broad availability of AI tools, in particular – to profit from new technological opportunities, and to respond to the needs of our civilisation, identified within a deeper humanistic reflection over its development.

The article is structured as follows:
− in Section 2, the state of the art in applications of AI in higher education is characterised;
− in Section 3, mathematical modelling as meta-concept of teaching MS&T is presented;
− in Section 4, psychosociological and humanistic aspects of teaching MS&T are highlighted;
− in Section 5, conclusions concerning an emerging paradigm of teaching MS&T are summarised.

This article is a significantly extended and updated version of the author's paper "Teaching measurement science and technology in the times of pervasive AI", presented at the 2024 IMEKO Congress in the session organised by the IMEKO Technical Committee TC1: Education and Training in Measurement and Instrumentation [34].

## 2. Application of AI in academic education

In the 2024 article [35], the applications of AI in higher education are grouped into four categories, *viz.* profiling and prediction, intelligent tutoring, assessment and evaluation, and adaptive personalisation:

- *Profiling and prediction* is focused on employing data-driven approaches to make informed decisions regarding students' academic pathways. It includes the use of AI tools for optimising admission-related decisions and course scheduling, for predicting and improving dropout and retention rates, as well as for recognising patterns in students' data.
- *Intelligent tutoring* includes the use AI tools for providing customer-fit instructional interventions, in particular – for teaching course content, for diagnosing students' strengths and weaknesses and offering personalised feedback, for developing appropriate learning materials, as well as for providing insights from the teachers perspective, useful for improving pedagogical strategies.
- *Assessment and evaluation* comprises the use of algorithms for automated grading, providing immediate and personalised feedback to students, evaluating student understanding and engagement, as well as implementing mechanisms for the evaluation of teaching methodologies and effectiveness.
- *Adaptive personalisation* consists in using AI tools for fitting educational experience to the individual needs of students. It includes shaping course contents, recommending personalised content and learning pathways, utilising academic data to monitor, guide, and support students.

The following journals, by virtue of their mission, are most predestined to publish articles devoted to the use of AI tools in higher education: *Active Learning and Higher Education*, *Assessment & Evaluation in Higher Education*, *Internet and Higher Education*, *Higher Education*, *Higher Education Research and Development*, *Journal of Higher Education*, *Research in Higher Education*, *Review of Higher Education*, *Studies in Higher Education*, and *Teaching in Higher Education*. Many important contributions to the topic under consideration appear, however, elsewhere – in the journals of more technical or interdisciplinary nature. Syntheses of practical guidelines and recommendations concerning the use of AI tools in education may be found in [36] and [37].

The development of AI has been an issue of philosophical, especially ethical, debate for more than 80 years. The implementation of AI tools in education has enhanced this debate with a didactic and psychosociological perspective [38–41]. During the recent two decades, numerous attempts have been made to synthetise the multidimensional experience of using AI tools in education. Here is a package of review papers, published in the 2020s, which characterise the state of the art in this domain: [42–54] and six papers on pages 69–172 of the book [55].

The protagonists of using AI tools in education point out that these tools (1) can help increase students' engagement by enabling them a more personalised and interactive learning experience, (2) may improve learning outcomes by providing them with access to personalised tutoring and immediate feedback, (3) can reduce teachers' workload by automating tasks such as grading home works [56]. What are the main concerns about using AI tools in education? Roughly speaking, they are the same as those related to other uses of AI tools, *viz.*: possible infringements of privacy, new opportunities for dishonesty, information uncertainty and disinformation; there is, moreover, a concern that overreliance on AI tools can reduce human interaction between teachers and students, and – consequently – hinder the social maturing of the latter [57].

By many academic teachers, AI tools are still perceived as the enemies of education: these tools are blamed for eroding critical thinking, enabling plagiarism, and displacing academic

staff – *cf.* [58]. It is true that, when deployed without ethical foresight or pedagogical reflection, AI tools can be misused, but we should not assume that such academic values as curiosity, rigour and creativity cannot coexist with AI systems. We should not blame the latter for generating new unsolvable problems, but rather notice that they reveal weak points of our routine approach to higher education, such as the bureaucratisation of learning, the ritualisation of assessment, and the widening gap between curricula and real life experience. In response to this challenge, we should try to reconsider the assumptions our educational systems are based upon; in particular – to redefine such basic concepts as knowledge, thinking and teaching in a new technological context; without going through pertinent philosophical reflection, we will not be able to reasonably decide whether students should be allowed to use AI tools in class, and to adequately redefine the role of an academic teacher – *cf.* [59]. Institutions of higher education must reclaim the future – they must preserve the moral and intellectual purpose of education, but not necessarily cling to traditional approaches and solutions at any price. Since MS&T is an interdiscipline with enormous diversity of applications, ranging from sciences to arts, it is an appropriate area for pioneering the implementation of this teaching philosophy.

Among the inevitable problems related to the above-suggested implementation, the issue of plagiarism requires deeper consideration here despite its universal nature. Noam Chomsky, the doyen of world linguistics, formulated it in a very concise and emphatic way: "*ChatGPT is basically high-tech plagiarism*" [60]. In fact, the training process, including both pretraining and fine-tuning, for AI tools like ChatGPT involves learning from a vast amount of both publicly available and licensed data, which enables those tools to generate responses based on patterns in those data. As a rule, however, the AI tools do not directly plagiarise existing contents – they generate new responses based on learned patterns rather than exactly copy phrases from original sources. Today, many artists and intellectuals use AI tools to brainstorm ideas, refine concepts, or stimulate their creativity. What about students? There are two extreme approaches to answering this question, and the whole spectrum of more moderate approaches logically falling between them. One extreme approach is to entirely ban the use of AI tools in class-works and home-works, the second – to tolerate such practice without any constrains. None of them can ignore the facts of academic life concerning actual use of AI tools by students. In December 2024, the Higher Education Policy Institute (London, UK) surveyed over 1 000 full-time undergraduate students of British institutions of higher education about the use of generative AI tools. The key finding reported is that 92% of them use such tools for various purposes: generating text (64%), enhancing and editing text (39%), summarising and note-taking (36%), translation or language support (35%), speech-to-text transcription (24%), generating images, videos or audio (19%), data analysis and presentation (15%) or writing computer code (15%) [61]. It is clear that any attempt to constrain this kind of students activities would be doomed to failure. A rational and constructive reply of academic staff and institutions should be rather an offer of diversified forms of education and training oriented on teaching safe and efficient use of AI tools without violating legal and ethical norms.

A global debate on the use of AI tools in academic education seams to give a slight preference to an intermediate solution based on somewhat nuanced ethical evaluation of the role of an AI tool in creative processes, *viz.* on the answer to the question whether this tool plays the role of an assistant to students or it replaces their genuine creativity. Of course, this demarcation may seem, very fuzzy, but it can be a good starting point for formulating more specific recommendations, in particular – the requirements concerning the role of an AI tool as an assistant. A tentative characterisation of that role could be as follows. A tool under consideration is an assistant if:
− it helps students to generate new ideas by providing inspiration, but not replacing them in making final decisions;

- it assists them in polishing a draft of a document, improving its logical coherence and suggesting editorial corrections, but not deciding about its core contents;
- it is a helpful collaborator providing "raw materials" and suggesting options of their processing, but never dictating the final shape of an intellectual product.

The ethical evaluation of an AI-aided work is always less problematic if its author follows the rule that every "borrowed" element of creative work is accompanied by a reference to its original source. This expectation may be met by appropriate prompting – a skill that should be developed by students under wise supervision of academic teachers. In this respect the AI tool called *Perplexity* (*https://www.perplexity.ai/*) is quite exceptional because, by principle, it draws on reliable sources from the internet and cites them for verification by its users [62].

Numerous methodologies of supporting educational processes with AI tools, overviewed in the literature referred to in this section, are potentially useful in teaching MS&T. Any adaptation of those methodologies to specific needs of teaching MS&T should, however, highlight ethical aspects of AI implementation – *cf.*, for example, [63–68] – and serve enhancing students' skills of critical thinking – *cf.*, for example, [69] and [70]. This recommendation is valid not only for the proponents of new courses in MS&T, but also for the academic teachers involved in the development of new curricula in other domains of technoscience. It will be, therefore, not considered here in more detail. The next section will focus on specific aspects of designing MS&T courses.

### 3. Mathematical modelling as meta-concept of teaching MS&T

We – teachers of measurement science and technology – need some deep reflection on the educational implications of accelerated penetration of AI tools to all the domains of technoscience and socio-economic life. The current practice of teaching MS&T is still predominantly based on a traditional (naïve) understanding of measurement, guided by the following premises:
- measurement is the only reliable source of information about material reality;
- mathematical modelling of material entities, being the essence of technoscience, entirely relies on measurement;
- providing data for mathematical modelling is just one of the numerous applications of measurement;
- evaluation of measurement uncertainty consists in comparing the actual result of measurement with universally agreed etalons or standards.

There are still in use the textbooks on MS&T, such as [71], which follow approaches of the 1960s and 1970s focused on extensive catalogues of MS&T components, sometimes quite well classified. Such approaches were criticised by Ludwik Finkelstein already in 1983: he pointed out that teaching MS&T should be based on general principles and logical reasoning rather than on the catalogues of ready for use solutions [72]. Although, the search for unifying concepts and principles of MS&T was undertaken in the 20th century, the first systematic elaboration of this topic in the form of a monograph appeared only in 2021 [73].

There is no doubt that in the age of pervasive AI, we should put more emphasis on concepts, principles, logical reasoning, and critical thinking. Let us support this statement with an example. The three-dimensional taxonomy of measurement sensors includes their categorisation according to (1) the measured quantity, (2) the operating principle, and (3) the application. The first category contains at least 14 types of sensors (acceleration sensors, flow sensors, gas sensors, humidity sensors, level sensors, light sensors, magnetic-field sensors, nuclear sensors, position sensors, pressure sensors, proximity sensors, sound sensors, strain sensors, and temperature sensors). The second category contains at least 7 kinds of sensors (capacitive sensors, inductive sensors, magnetic sensors, optical sensors, piezoelectric sensors, resistive sensors, and ultrasonic sensors). The third category contains at least 7 groups of sensors (aerospace sensors, automotive sensors, biomedical sensors, consumer electronics sensors, environmental sensors, industrial sensors, and military sensors). Of course, the students of MS&T should be aware of this structured taxonomy of sensors, but not necessarily learn exhaustive specification of each type, each kind and each group of sensors. They will be able to autonomously acquire and understand the exhaustive specification of any type, any kind or any group of sensors, provided they are equipped with adequate patterns of their characterisation based on their mathematical models. This is just an example, but a very important one since sensors are the most distinctive parts of measuring systems. Convergent technologies contribute to relatively quick unification and standardisation of the other constitutive elements such systems, while the progress of unification and standardisation of sensors is much slower.

The protagonists of the traditional understanding of measurement, as a rule, ignore the dependence of measurement on mathematical models, more precisely – the roles of those models in defining measurands, calibrating measurement channels, estimating measurands and evaluating measurement uncertainty, although – as already mentioned in Section 1 – basic tools of AI, *viz.* universal approximators (such as neural networks), have been already for years used in modelling measurement channels, and consequently – in performing those operations.

Thinking in terms of semantic and mathematical models of material entities (phenomena, objects or processes) is a natural way of dealing with the problems of MS&T in the age of ubiquitous use of AI tools which are, as a rule, semantic or mathematical models: large language models or universal approximators. Mathematical modelling of a material entity is always preceded by its semantic modelling. The latter is an operation of fundamental importance from an epistemic point of view because it consists in crossing the boundary between the sphere of material entities and the sphere of their abstract images. Since mathematical modelling is indispensable for defining measurands, designing measuring systems and evaluating measurement uncertainty, it may be considered a meta-concept for building educational frameworks in the field of MS&T. Detailed descriptions of the instances of mathematical modelling in MS&T should be included in the corresponding undergraduate courses, and a general methodology of mathematical modelling – in the relevant graduate courses.

The meta-concept of mathematical modelling is closely related to another meta-concept of key importance, *viz. understanding* – in the sense of our ability to recognise the properties of a material entity and to use the knowledge thus obtained for solving related practical problems. The latter clarification is important because the word *understanding* may be ambiguous: its meaning significantly depends on the context [74], and is a subject of deep philosophical considerations – *cf.* [75] or [76]. In teaching MS&T, its interpretation as the ability to recognise causal relationships and to use the knowledge thus obtained for solving measurement-related problems of practical nature is satisfactory in the majority of cases; some weakly defined measurements [77], however, cannot be properly understood under this constraint imposed on the definition of understanding.

The art of modelling causal relationships is of key importance both for the methodology of designing measuring systems and for making measurement-based decisions of ethical nature (to be addressed in Section 4). The existence of cause-effect relationships in material reality cannot be proven, neither by logical nor by experimental means. However, taking into account the practical productivity of reasoning guided by such relationships, we accept a paradigm of causality and make endeavours to built semantic and mathematical models of material entities, which are assumed to represent those relationships. Discovering causal relations and making use of them is not only a distinctive feature of understanding but also of human intelligence in general. It should be noted that there is no closed-form definition of causality that would not provoke any logical or philosophical objections. There is, however, a vast practical experience concerning identification of simple networks of causal relationships, especially – the networks without loops. The currently available AI tools, although very useful as human assistants in this respect, fail to meet our expectations driven by the needs related to engineering design (in particular – the design of measuring systems and devices) and to the ethical decision-making (in particular – those related to measurement applications). Judea Pearl, an Israeli-American computer scientist and philosopher who in the 1980s pioneered the development of causal models in AI in the form of so-called Bayesian networks, concluded in 2018 that future AI systems must incorporate causal models to reach the level of human-like intelligence [78]. It seems that achieving this goal is harder than expected; therefore, causality continues to be an active and important topic in AI-related research. The identification of multidimensional causal networks with multiple loops is a challenge that should be urgently undertaken for the sake of safe use of AI tools, especially those based on large language models, in applications directly concerning human lives and wellbeing – now and in the future. The functioning of large language models is based on correlation-type relationships among linguistic items; the supervision of their operation based on deductive reasoning and causal relationships is needed to make this functioning safe and reliable. This is why teaching students of MS&T the art of causal modelling is so critically indispensable.

A general methodology for mathematical modelling of measurement-related entities is systematically presented in Chapter 6 of the author's handbook *Technoscientific Research: Methodological and Ethical Aspects* [79]. It is a refined version of the methodology proposed by the author in 2013 [80] – the version that takes into account his later contributions, *viz.* the articles [81] and [82]. A meta-model of measurement described there includes generic mathematical models indispensable for defining measurands, calibrating measurement channels, and providing measurement results in the form of measurands' estimates accompanied by some indicators of their uncertainty. Each of those models is to be concretised in any specific case of measurement. Of course, the AI tools, such as universal approximators, can be used for this purpose. However, there are some limitations in their more-or-less routine application, *viz.* limitations due to the fact that measurand estimation and uncertainty evaluation are operations of abductive nature. Unlike deductive reasoning, abductive reasoning is highly uncertain, and its uncertainty cannot be mitigated solely by increasing the volume of data (like in the case of inductive reasoning), but additional information about the material (physical, chemical, biological…) background of measurement under consideration is necessary to sufficiently constrain the set of admissible estimates of a measurand [83]. Unfortunately, abductive reasoning is an underdeveloped aspect of the today's AI tools which are predominantly based on inductive reasoning, *i.e.* on data-voracious machine learning, and which are not creative enough to solve non-trivial problems of abduction – as argued in Part II of the book [84]. The advancement in this area is a *sine qua non* condition of the full development of explainable AI, and the progress in the latter area is a *sine qua non* condition of the progress in the development of trustworthy AI, indispensable for predictable and controllable expansion of global-scale applications of AI.

For the time being, there is a lot of research activities oriented on the advancement of explainable AI. Enough to say that a 111-page article [85], providing a systematic review of the literature on applications of explainable AI, refers to 512 articles which were published in peer-reviewed journals only in the years 2021–2023. This is a fact of importance for the development of MS&T curricula, since the use of the tools of explainable AI for designing measuring systems, both in the role of the design tools and their elements, enables the reliable evaluation of the AI-generated quantitative components of measurement uncertainty. However, these tools may occur insufficient for assessing total measurement uncertainty, that includes also fundamental epistemic uncertainty, as this would require taking into account some qualitative

– mainly psychosociological and humanistic – aspects of AI development towards trustworthy AI. Related topics should be urgently incorporated in MS&T curricula.

One more argument in favour of enhancing the intellectual toolbox of MS&T graduates with the skills related to abductive reasoning follows form the realisation of the importance of this type of reasoning in technoscience and everyday life: it is the basis of any non-trivial measurement [86], of scientific explanation, of medical and technical diagnostics, of criminal investigation, and of reliable communication between two human beings. Its creative potential cannot be fully developed without combining complementary strong features of human and artificial intelligence. The following example is quite indicative in this respect. In a study conducted at Stanford University and reported in the 2024 article [87], an AI system used as a diagnostic tool outperformed a group of 50 physicians. It turned out, however, that the availability of that system to the latter did not significantly improve their clinical reasoning. This finding indicates the need for significant advancement of research on the mode of "collaboration" between human and artificial intelligence, which is impossible without fusion of technical and non-technical competences in the engineering curricula, in particular – in the MS&T-related curricula.

## 4. Psychosociological and humanistic aspects of teaching MS&T

The situation outlined in the previous section may engender problems of ethical nature that were identified already 80 years ago, but have been generally recognised only recently in the context of the global discussion over intelligent text robots such as ChatGPT. In particular, this discussion has exposed some problems less realised in the past, *viz.* the ones related to legal protection of intellectual property. The 2024 Nobel Prize in Chemistry is an excellent example showing how insightful the Noam Chomsky's remark (quoted in Section 2) is since that prize was awarded for the computational design of proteins and the prediction of their structure, aided by an AI tool called AlphaFold2. From the ethical perspective, that decision may be perceived as problematic because the awarded achievement is based on an AI-model trained on the results of long-run research of numerous scientists all over the world – not just on the research contributions of the awardees.

Many other ethical problems may directly affect MS&T – for example, when measuring instrumentation is applied in biomedical engineering. In the latter case, the evaluation of uncertainty must be precise enough to exclude the risk of harmful measurement-based decisions concerning human health or wellbeing. Less evident may seem the risks related to unreliable measurements embedded in such technical objects as autonomous vehicles, social robots or industrial control systems.

Over the 25 centuries of development of Western philosophy, many proposals for systematically solving moral problems have appeared. Fortunately, their deeper analysis leads to the conclusion that each of them can be viewed as a combination of three monistic ethics:
− *virtue ethics* which considers an act to be morally good if it is performed by a virtuous person;
− *deontological ethics* which considers an act to be morally good if it constitutes the fulfilment of a duty or law;
− *consequentialist ethics* which considers an act to be morally good if its consequences are good.

The importance of the above components differs from one ethical system to another, but regardless of their proportions a universal methodology of making morally significant decisions may be formulated – the methodology inspired by the practice of making engineering decisions based on mono-criterial optimisation. The key for its formulation is the analogy linking:
− the consequences of our actions with optimisation criteria;

- the relevant moral norms with optimisation constraints;
- the virtues of decision-makers with the numerical advantages of optimisation algorithms.

Using this analogy, we may postulate that – when faced with the need to undertake a morally significant decision – one should make the best choice in terms of the anticipated consequences, but only within the space of admissible solutions, defined by relevant moral norms. The practical implementation of this pattern of decision-making requires the use of causal models for prediction of consequences of an act and a catalogue of norms generating constraints to be satisfied by that act. Numerous attempts to compile such a catalogue have been made – at least since 1942 when Isaac Asimov, in a short story published in *Astounding Science Fiction* magazine, proposed three laws of robot ethics. Over last eighty years, numerous organisations have proposed various sets of principles for ethical AI [88]; among them, a set of 23 recommendations for developers of AI systems in a document launched during the AI conference organised in 2017 in Asilomar (California, USA) by the Future of Life Institute, and called "Asilomar AI Principles" [89]. The Canadian philosopher of science and medicine, Paul R. Thagard, has demonstrated [90] that they may be logically derived from four principles of bioethics, introduced in 1978 by the American ethicists Thomas L. Beauchamp and James F. Childress [91], *viz.*: (1) the principle of beneficence, which obliges us to maximise the health benefits of patients; (2) the principle of non-maleficence, which obliges us to minimise the harm associated with experimental activities with their involvement; (3) the principle of respect for autonomy, which obliges us to respect patients' decisions regarding their own lives; (4) the principle of justice, which obliges us to give patients everything they deserve, to treat them equally, fairly and impartially.

The main advantage of the above-outlined methodology of solving morally significant problems in AI-aided engineering, in particular – in engineering related to MS&T – consists in its high potential for automatic implementation. Moreover, this methodology is helpful in integrating ethical issues with technoscientific issues in the areas of morally sensitive applications of measurement, such as social robotics or biomedical monitoring. It enables a productive collaboration of engineers with professional ethicists, based on a common language resulting from the above-proposed analogy between constrained optimisation and making morally significant decisions.

The recognition of the need to include ethical considerations in engineering education seems to be a harbinger of the imminent recognition of the need to significantly enhance technoscientific curricula with psychosociological and humanistic contents – to overcome a divide between sciences and humanities that appeared in the 19th century and deepened in the 20th century as a result of overspecialisation. Already in the late 1960s, Charles P. Snow, in his lecture *The Two Cultures and the Scientific Revolution* [92], formulated the thesis about the incompatibility of two cultures having common historical roots in Western philosophy, *viz.* sciences and humanities. He argued that the representatives of these cultures had ceased to understand each other, and warned about the dangerous consequences of the deepening intellectual gap between them. Over the past six decades, Charles P. Snow's thesis has been the subject of critical discussion, which has made many thinkers aware of the need to reintegrate sciences and humanities. However, it was not until the age of ubiquitous use of AI tools that awareness of this need became common. The progressing convergence of technoscientific disciplines critically dependent on IT tools (including AI tools) requires an adequate convergence of technoscientific and humanistic education at all its stages – the introduction of selected elements of humanistic knowledge to technoscientific curricula and selected elements of technoscientific knowledge to educational programmes in humanities. Sustainable development and implementation of modern technologies is impossible without close interdisciplinary collaboration which involves not only experts in various domains of engineering, but also representatives of social sciences (especially psychologists and sociologists) and humanities (especially ethicists and philosophers of technoscience). In order to effectively communicate, they must master a kind of common language. This is only a very brief *ad hoc* justification of the need to reintegrate technoscience and humanities in academic education, but there are also some deeper reasons (of philosophical nature) supporting this idea.

The essence of the unity of sciences and humanities has been accurately grasped on the portal *ScienceFather*: a synergy resulting from the integration of humanities and sciences implies the fusion of the analytical rigor of scientific inquiry with the nuanced insights of humanistic inquiry, and in a holistic understanding of the world encompassing both empirical knowledge and human experience [93]. Albert Einstein justified this essence by referring to the Western intellectual tradition. According to him "*All religions, arts and sciences are branches of the same tree*": their aim is to spread moral and cultural understanding without referring to force [94]. Further justification may be found in the preface to the 2018 report *The Integration of the Humanities and Arts with Sciences, Engineering, and Medicine in Higher Education*, where its authors recommend – despite limited causal evidence on the impact of integration on students – that further effort be made to develop and disseminate a variety of approaches to integrated education and to carry out research on the impact of such forms of education on students [95].

Examples of partial integration of technoscience and humanities in academic education may be already found in many Western institutions of higher education. Here are selected examples:
− Lehigh University (Bethlehem, Pennsylvania, USA) offers an undergraduate programme leading to the Integrated Degree in Engineering, Arts and Sciences. It is dedicated to students with high creative potential – future innovators able to use different styles of thinking ("*renaissance thinking for the technological era*") [96].
− The graduate programmes in Human-centred Artificial Intelligence are offered by several European universities (Göteborgs Universitet [97], Vrije Universiteit Amsterdam [98], Università degli studi di Milano [99], as well as by the consortium including Technological University Dublin, Università degli Studi di Napoli Federico II, De Hogeschool Utrecht, Budapesti Műszaki és Gazdaságtudományi Egyetem [100]); the corresponding curricula provide students with a deep understanding of how AI is transforming society, as well as with a broad theoretical preparation for analysing the consequences of the proliferation of AI tools.

Such educational offers are multiplying as IT employers increasingly value interdisciplinary competences in their employees, such as the ability to clearly communicate thoughts (both in written and oral form), productivity in teamwork, the ability to make morally significant decisions and the ability to apply knowledge for solving complex interdisciplinary problems – *cf.* [101]. They expect that graduates of engineering studies will not only be able to take up employment in a profession corresponding to the profile of their education, but also will manage to adapt their competences to changing needs in the following years of their professional careers – *cf.* [102].

According to the report about the global labour market, published in 2023 under the auspices of the World Economic Forum, over 60% of surveyed employers indicate the growing importance of the following competencies: *creative thinking*, *analytical thinking*, *technological literacy*, *curiosity and lifelong learning* and the broadly understood ability to respond flexibly (*resilience, flexibility and agility*) – *cf.* [103]. It seems that in today's approach to technoscientific education, including teaching of MS&T, we are inclined to overvalue purely technical skills and undervalue skills from the other four categories – the skills relying on the art of asking questions.

Why is asking questions so important today? In times of gigantic databases and intelligent search engines, in times of ubiquitous sensor networks – collecting measurement data that may be useful to someone somewhere-sometime – the ability to extract the useful and reliable information from these resources and to assess its credibility is becoming a key competence. Asking good questions is a form of creativity, the value of which we appreciate when using AI tools such as ChatGPT. The socio-economic importance of new professionals, called *prompters*, is growing rapidly. As a rule, holders of diplomas in humanities are better candidates for this profession than engineers, but this observation should not imply a conclusion that an engineer cannot acquire prompter qualifications by attending classes on logic, rhetoric, language culture or propaedeutics of philosophy.

Being aware that modern economies are based on knowledge, we must recognise that: (1) verified (or warranted) information and the methodology for obtaining it are crucial for the lifelong professional career of an engineer; (2) mastering this methodology is more important than learning detailed specialist contents that will, in practice, become outdated within three to five years after their appearance. The ability to learn independently and to define learning goals independently – these are the main competences of the future, with which we should equip our graduates. The implementation of this postulate requires very careful selection of the contents to be memorised – a selection that gives priority to structural and methodological knowledge because this type of knowledge plays a fundamental role in finding and verifying information on any topic, and in particular – in asking questions skilfully.

In the 1990s, efficient use of computers and measurement instrumentation was impossible without certain hardware- and software-related skills. Today, such competences are not needed, user manuals have become obsolete; we expect that, when carrying out a specific task, we will communicate with a measuring system via an intelligent multimedia-based interface enabling us not only to operate this system, but also to compensate for our lack of precision in operation, to suggest a spectrum of alternative ways of operation, *etc*. Designing such an interface requires creative cooperation between an instrumentation engineer and a person whose competences include verbal and iconic communication, psychology of human needs and reflexes, as well as a methodology of evaluating new technologies from the point of view of ethics and ergonomics.

## 5. Conclusions

The conclusions presented in this section are intended to integrate key elements of the emerging paradigm of MS&T teaching announced in the title of the article – the elements that have been introduced and analysed in the previous sections. Many of them are not specific to MS&T, and can be applied to teaching students of various subfields of IT – and even of those subfields of engineering which are not focused on IT. This is one more illustration of the phenomenon called *convergence of technologies*.

Taking into account the omnipresence and key role of measurement in industry, medicine, economy and other sectors of social practice, we – teachers of MS&T – should recognise the pragmatic and ethical implications of the evolving technological and socio-economic environment of educational practice. The corresponding evolution of the engineering curricula (rather than of isolated courses devoted to MS&T) should consist in inclusion of the following contents:
− general methodology of mathematical modelling of material entities;
− procedures of effective use of mathematical models in conceptual and functional design of measuring systems;
− strategies of responsible use of AI tools for this purpose;
− elements of research-and-engineering ethics, enabling the dialogue between engineers and professional ethicists over AI-induced moral issues.

The scope and complexity of the above-listed contents should be fit to the level of instruction: they should be more basic and illustrative at the B.Sc. level, and significantly more advanced and abstract at the Ph.D. level. The range and strategy of introducing postulated changes should also be fit to the local educational environment. A considerable inertia of curricula evolution is characteristic of most academic institution, but this cannot be used as a justification for procrastination. The appropriate change of selected syllabi, which – as a rule – is easier from institutional perspective, may be a good anticipatory step preparing the ground for reforming an entire curriculum.

If the use of mathematical models is concerned, the meta-model of measurement, referred to in Section 3, could become a convenient platform for incorporation of the methodology of mathematical modelling into MS&T courses. If the ethical contents are concerned, Chapters 11–20 of the textbook [104] could be a good starting point even at the level of B.Sc. studies. The first-semester students of cybersecurity and the students of internet of things, exposed to such contents at the author's institution, are able to grasp them; so, it is reasonable to expect that MS&T students at the B.Sc. level will be able to follow their suite.

Over the past decades, we have come to realise that a very limited share of MS&T graduates are going to effectively use knowledge and skills concerning the physical phenomena underlying functioning of sensors and other hardware components of measurement instrumentation, while a great many of them will need some competencies concerning methodology of mathematical modelling and data processing, in particular – of using AI-based algorithms. It would be, however, a mistake to deprive the MS&T students of any opportunity to gain understanding of those phenomena – not only for philosophical reasons, but also for the need to prepare them to effectively communicate, over their professional careers, with the specialists of empirical sciences.

We have already learned that many operations related to the design and utilisation of measuring instrumentation may be ceded to AI systems, in particular, such operations as searching databases of technical information about sensors and other hardware components of measuring systems, optimal fitting of those components to the design requirements, automatic calibration of measurement channels, automatic generation of technical documentation and its translation in various languages, *etc*. Consequently, the skills related to the efficient use of AI tools are getting increasingly more important than memorising catalogues of hardware and software components of measuring instrumentation; the ability to use general methodologies of designing technical objects and general principles of designing measuring systems are getting increasingly more important than technical data of measurement bridges or structures of analogue-to-digital and digital-to-analogue converters. We – teachers of measurement science and technology – should realise that "*Academia needs to adapt to a more AI-integrated system* [since] *AI assistance is not a temporary trend, but a tendency of technological progress, which should be treated with cautious openness*" [105]; that already today advanced AI tools contribute to the most creative activities of technoscientific research – *cf.*, for example, [106]. Several advanced AI engineering laboratories all over the world are working on developing AI tools intended to support the research process carried out according to the so-called *scientific method*, the essence of which is generating hypotheses and their experimental verification. A good example is AI co-scientist – a multi-agent system designed as a collaborative tool for scientists [107]. The advanced, specialised AI tools do not eliminate general-purpose tools, like Microsoft Copilot, from everyday research practice; for example, Google Translator was used to optimise the wording of some paragraphs in this article.

## Acknowledgment

This work was supported by the Institute of Radioelectronics & Multimedia Technology at the Faculty of Electronics & Information Technology, Warsaw University of Technology, Poland [from the State block grants 2024 &2025].

**References**

[1] S. Legg, M. Hutter, "A collection of definitions of intelligence", 2007, *arXiv: 0706.3639* [2025-12-05].

[2] C. Ategeka, *The Unintended Consequences of Technology: Solutions, Breakthroughs, and the Restart We Need*, Wiley & Sons, Hoboken (USA) 2022, pp. 62–63.

[3] *EU Artificial Intelligence Act*, European Parliament, 2024, *https://artificialintelligenceact.eu/article/3/* [2025-12-05].

[4] L. Finkelstein, D. Hofmann, "Intelligent measurement – a view of the state of the art and current trends", *Measurement,* 1987, Vol. 5, No. 4, pp. 151–153.

[5] C. Alippi, A. Ferrero, V. Piuri, "Artificial intelligence for instruments and measurement applications", *IEEE Instrumentation & Measurement Magazine,* 1998, Vol. 1, No. 2, pp. 9–17.

[6] F. Amigoni, A. Brandolini, G. D'Antona, R. Ottoboni, M. Somalvico, "Artificial intelligence in science of measurements: From measurement instruments to perceptive agencies", *IEEE Transactions on Instrumentation and Measurement,* 2003, Vol. 52, No. 3, pp. 716–723.

[7] Chao Liu, Jiashu Sun, "AI in Measurement Science", *Annual Review of Analytical Chemistry,* 2021, Vol. 14, No. 1, pp. 1–19.

[8] M. Khanafer, S. Shirmohammadi, "Applied AI in instrumentation and measurement: The deep learning revolution", *IEEE Instrumentation & Measurement Magazine,* 2020, Vol. 23, No. 6, pp. 10–17.

[9] V. Polužanski, U. Kovacevic, N. Bacanin, T. A. Rashid, S. Stojanovic, B. Nikolic, "Application of machine learning to express measurement uncertainty", *Applied Sciences,* 2022, Vol. 12, No. 17, pp. 1–13 of Article #8581.

[10] A. Rashed, S. Shirmohammadi, "A Novel Method to Estimate Measurement Error in AI-Assisted Measurements", *Proc. 2022 IEEE International Instrumentation and Measurement Technology Conference (I2MTC)*, pp. 1–5, *https://ieeexplore.ieee.org/document/9806449* [2025-12-05].

[11] A. Ahmed, S. Aziz, A. Abd-Alrazaq, F. Farooq, M. Househ, J. Sheikh, "The Effectiveness of Wearable Devices Using Artificial Intelligence for Blood Glucose Level Forecasting or Prediction: Systematic Review", *Journal of Medical Internet Research,* 2023, Vol. 25, pp. 1–16 of Article #e40259.

[12] A. Al Maashri, A. Saleem, H. Bourdoucen, O. Eldirdiry, A. Al Ghadani, "A novel drone-based system for accurate human temperature measurement and disease symptoms detection using thermography and AI", *Remote Sensing Applications: Society and Environment,* 2022, Vol. 27, pp. 1–15 of Article #100787.

[13] I. Hamelink, E. de Heide, G. J. Pelgrim, T. C. Kwee, P. M. A. van Ooijen, G. H. de Bock, R. Vliegenthart, "Validation of an AI-based algorithm for measurement of the thoracic aortic diameter in low-dose chest CT", *European Journal of Radiology,* 2023, Vol. 167, pp. 1–7 of Article #111067.

[14] Han Bao, Kejia Zhang, Chenhao Yu, Hu Li, Dan Cao, Huazhong Shu, Luwei Liu, Bin Yan, "Evaluating the accuracy of automated cephalometric analysis based on artificial intelligence", *BMC Oral Health,* 2023, Vol. 23, No. 1, pp. 1–10.

[15] Kai Huang, *et al.* [31 names], "Artificial Intelligence–Based Psoriasis Severity Assessment: Real-world Study and Application", *Journal of Medical Internet Research,* 2023, Vol. 25, pp. 1–14 of Article #e44932.

[16] Ming-Wun Wong, B. D. Rogers, Min-Xiang Liu, Wei-Yi Lei, Tso-Tsai Liu, Chih-Hsun Yi, Jui-Sheng Hung, Shu-Wei Liang, Chiu-Wang Tseng, Jen-Hung Wang, Ping-An Wu, Chien-Lin Chen, "Application of Artificial Intelligence in Measuring Novel pH-

Impedance Metrics for Optimal Diagnosis of GERD", *Diagnostics,* 2023, Vol. 13, No. 5, pp. 1–13 of Article #960.

[17] J. Rade, J. Zhang, S. Sarkar, A. Krishnamurthy, J. Ren, A. Sarkar, "AI Guided Measurement of Live Cells Using AFM", *IFAC-PapersOnLine,* 2021, Vol. 54, No. 20, pp. 316–321.

[18] D. Reifs, L. Casanova-Lozano, R. Reig-Bolaño, S. Grau-Carrion, "Clinical validation of computer vision and artificial intelligence algorithms for wound measurement and tissue classification in wound care", *Informatics in Medicine Unlocked,* 2023, Vol. 37, pp. 1–12 of Article #101185.

[19] Shu-Yin Chiang, Yi-Feng Chen, "Contactless Blood Pressure Measurement by AI Robot", *Sensors & Materials,* 2022, Vol. 34, No. 11, pp. 4167–4184.

[20] J. W. van der Graaf, M. L. van Hooff, B. van Ginneken, M. Huisman, M. Rutten, D. Lamers, N. Lessmann, M. de Kleuver, "Development and validation of AI-based automatic measurement of coronal Cobb angles in degenerative scoliosis using sagittal lumbar MRI", *European Radiology,* 2024, pp. 1–10, *https://doi.org/10.1007/s00330-024-10616-8 [2025-12-05].*

[21] S. Vogt, C. Scholl, P. Grover, J. Marks, M. Dreischarf, U.-D. Braumann, P. Strube, A. Hölzl, S. Böhle, "Novel AI-based algorithm for the automated measurement of cervical sagittal balance parameters. A validation study on pre-and postoperative radiographs of 129 patients", *Global Spine Journal,* 2025, Vol. 15, No. 2, pp. 1155–1165.

[22] Yihan Zhang, Yubing Hu, Nan Jiang, A. K. Yetisen, "Wearable artificial intelligence biosensor networks", *Biosensors and Bioelectronics,* 2023, Vol. 219, pp. 1–17 of Article #114825.

[23] Chenang Liu, Rongxuan Wang, Ian Ho, Z. Kong, C. Williams, S. Babu, C. Joslin, "Toward online layer-wise surface morphology measurement in additive manufacturing using a deep learning-based approach", *Journal of Intelligent Manufacturing,* 2023, Vol. 34, No. 6, pp. 2673–2689.

[24] T. Deleruyelle, A. Auguste, F. Sananes, G. Oudinet, "Nondestructive Diagnostic Measurement Methods for HF RFID Devices With AI Assistance", *IEEE Open Journal of Instrumentation and Measurement,* 2023, Vol. 2, pp. 1–10.

[25] D. Y. Pimenov, A. Bustillo, S. Wojciechowski, V. S. Sharma, M. K. Gupta, M. Kuntoğlu, "Artificial intelligence systems for tool condition monitoring in machining: Analysis and critical review", *Journal of Intelligent Manufacturing,* 2023, Vol. 34, No. 5, pp. 2079–2121.

[26] Zhiqin Zhu, Yangbo Lei, Guanqiu Qi, Yi Chai, N. Mazur, Yiyao An, Xinghua Huang, "A review of the application of deep learning in intelligent fault diagnosis of rotating machinery", *Measurement,* 2023, Vol. 206, pp. 1–24 of Article #112346.

[27] M. El Ghadoui, A. Mouchtachi, R. Majdoul, "Intelligent surface roughness measurement using deep learning and computer vision: a promising approach for manufacturing quality control", *The International Journal of Advanced Manufacturing Technology,* 2023, Vol. 129, No. 7, pp. 3261–3268.

[28] A. S. Mahammad, "Using AI in Dimensional Metrology", [in] *Handbook of Metrology and Applications* (Eds. D.K. Aswal, S. Yadav, T. Takatsuji, P. Rachakonda, H. Kumar), Springer Nature, Singapore 2023, pp. 1025–1042.

[29] Nie Shaojun, Wang Yunpeng, Zonglin Jiang, "Force measurement using strain-gauge balance in shock tunnel based on deep learning", *Chinese Journal of Aeronautics,* 2023, Vol. 36, No. 8, pp. 43–53.

[30] M. Wieczorowski, D. Kucharski, P. Sniatala, P. Pawlus, G. Krolczyk, B. Gapinski, "A novel approach to using artificial intelligence in coordinate metrology including nano scale", *Measurement,* 2023, Vol. 217, pp. 1–11 of Article #113051.

[31] Xi-Feng Liu, Hong-Hu Zhu, Bing Wu, Jie Li, Tian-Xiang Liu, Bin Shi, "Artificial intelligence-based fiber optic sensing for soil moisture measurement with different cover conditions", *Measurement,* 2023, Vol. 206, pp. 1–12 of Article #112312.

[32] T. Adel, S. Bilson, M. Levene, A. Thompson, "Trustworthy Artificial Intelligence in the Context of Metrology", 2024, *arXiv: 2406.10117* [2025-12-05].

[33] P. Ala-Pietilä, Y. Bonnet, U. Bergmann, M. Bielikova, C. Bonefeld-Dahl, W. Bauer, L. Bouarfa, R. Chatila, M. Coeckelbergh, V. Dignum, *The Assessment List for Trustworthy Artificial Intelligence (ALTAI)*, European Commission, Brussels 2020, *https://digital-strategy.ec.europa.eu/en/library/assessment-list-trustworthy-artificial-intelligence-altai-self-assessment* [2025-12-05].

[34] R. Z. Morawski, "Teaching measurement science and technology in the times of pervasive AI", *Measurement: Sensors,* 2025, Vol. 8, pp. e1–e4 of Article #101315, *https://doi.org/10.1016/j.measen.2024.101315* [2025-12-05].

[35] M. Bond, H. Khosravi, M. De Laat, N. Bergdahl, V. Negrea, E. Oxley, P. Pham, S. W. Chong, G. Siemens, "A meta systematic review of artificial intelligence in higher education: a call for increased ethics, collaboration, and rigour", *International Journal of Educational Technology in Higher Education,* 2024, Vol. 21, No. 1, pp. 1–41 of Article #44.

[36] UNESCO, "Artificial intelligence in education"*, 2022–2025, https://www.unesco.org/en/digital-education/artificial-intelligence?hub=32618* [2025-12-05].

[37] K. Swargiary, *Embracing AI in Education: A Guide for Teachers*, Lambert Academic Pub., London (UK) 2024.

[38] Bai Xuejiao, "The Role and Challenges of Artificial Intelligence in Information Technology Education", *Pacific International Journal,* 2024, Vol. 7, No. 1, pp. 86–92.

[39] S. Erduran, "How artificial intelligence is changing science education", *University World News,* February 10, 2024, *https://www.universityworldnews.com/post.php?story=20240206105516722* [2025-12-05].

[40] N. Matthijs, "What AI means for higher education teaching: hype vs reality", *University World News,* January 27, 2024, *https://www.universityworldnews.com/post.php?story=20240123095419557* [2025-12-05].

[41] S. Nemorin, A. Vlachidis, H. M. Ayerakwa, P. Andriotis, "AI hyped? A horizon scan of discourse on artificial intelligence in education (AIED) and development", *Learning, Media and Technology,* 2023, Vol. 48, No. 1, pp. 38–51.

[42] M. Bearman, J. Ryan, R. Ajjawi, "Discourses of artificial intelligence in higher education: A critical literature review", *Higher Education,* 2023, Vol. 86, No. 2, pp. 369–385.

[43] M. A. Chaudhry, E. Kazim, "Artificial Intelligence in Education (AIEd): A high-level academic and industry note 2021", *AI and Ethics,* 2022, Vol. 2, pp. 157–165.

[44] T. K. F. Chiu, Qi Xia, Xinyan Zhou, Ching Sing Chai, Miaoting Cheng, "Systematic literature review on opportunities, challenges, and future research recommendations of artificial intelligence in education", *Computers and Education: Artificial Intelligence,* 2023, Vol. 4, pp. 1–15 of Article #100118.

[45] H. Crompton, D. Burke, "Artificial intelligence in higher education: the state of the field", *International Journal of Educational Technology in Higher Education,* 2023, Vol. 20, No. 1, pp. 1–22.

[46] W. Holmes, I. Tuomi, "State of the art and practice in AI in education", *European Journal of Education,* 2022, Vol. 57, No. 4, pp. 542–570.

[47] Jiahui Huang, S. Saleh, Yufei Liu, "A review on artificial intelligence in education", *Academic Journal of Interdisciplinary Studies,* 2021, Vol. 10, No. 3, pp. 206–217.

[48] Lijia Chen, Pingping Chen, Zhijian Lin, "Artificial intelligence in education: A review", *IEEE Access,* 2020, Vol. 8, pp. 75264–75278.
[49] P. Limna, S. Jakwatanatham, S. Siripipattanakul, P. Kaewpuang, P. Sriboonruang, "A review of artificial intelligence (AI) in education during the digital era", *Advance Knowledge for Executives,* 2022, Vol. 1, No. 1, pp. 1–9.
[50] A. Marengo, A. Pagano, K. Soomro, J. Pange, "The educational value of artificial intelligence in higher education: a ten-year systematic literature review", *Interactive Technology and Smart Education,* 2024, Vol. 21, No. 1, pp. 1–18.
[51] J. Schürmann, M. Talmeier, "ChatGPT in der Lehre: Chancen und Risiken der KI-gestützten Sprachmodelle", Mittelstand-Digital Zentrum Berlin, *www.digitalzentrum-berlin.de* [2025-12-29].
[52] F. A. Triansyah, I. Muhammad, A. Rabuandika, K. D. Pratiwi, N. Teapon, M. S. Assabana, "Bibliometric Analysis: Artificial Intelligence (AI) in High School Education", *Jurnal Imiah Pendidikan Dan Pembelajaran,* 2023, Vol. 7, No. 1, pp. 112–123.
[53] Xieling Chen, Di Zou, Haoran Xie, Gary Cheng, Caixia Liu, "Two decades of artificial intelligence in education", *Educational Technology & Society,* 2022, Vol. 25, No. 1, pp. 28–47.
[54] Xuesong Zhai, Xiaoyan Chu, Ching Sing Chai, M. Siu Yung Jong, A. Istenic, M. Spector, Jia-Bao Liu, Jing Yuan, Yan Li, "A Review of Artificial Intelligence (AI) in Education from 2010 to 2020", *Complexity,* 2021, Vol. 2021, pp. 1–18 of Article #8812542.
[55] A. Chakir, J. F. Andry, A. Ullah, R. Bansal, M. Ghazouani, *Engineering Applications of Artificial Intelligence*, Springer Nature, Cham (Switzerland) 2024, 69–172.
[56] T. J. Sejnowski, *ChatGPT and the Future of AI: The Deep Language Revolution*, The MIT Press, Cambridge (USA) & London (UK) 2024, p. 28.
[57] *ibid*.
[58] J. Y. Auh, "Wrong battle? Universities must lead, not fight, AI integration", *University World News,* May 25, 2025, *https://www.universityworldnews.com/post.php?story=20250522152541838&utm_source=newsletter&utm_medium=email&utm_campaign=COMMNL8037* [2025-12-25].
[59] *ibid*.
[60] "Professor Noam Chomsky on ChatGPT and Education", [in] *YouTube* January 2023, *https://www.youtube.com/watch?v=SJi4VE-0MoA* [2025-12-05].
[61] J. Freeman, *Student generative AI survey 2025*, Higher Education Policy Institute, London 2025, *https://www.hepi.ac.uk/wp-content/uploads/2025/02/HEPI-Policy-Note-61.pdf* [2025-12-05].
[62] T. J. Sejnowski, *ChatGPT and the Future of AI: The Deep Language Revolution*, 2024, p. xi.
[63] S. Akgun, C. Greenhow, "Artificial intelligence in education: Addressing ethical challenges in K-12 settings", *AI and Ethics,* 2022, Vol. 2, No. 3, pp. 431–440.
[64] A. Aler Tubella, M. Mora-Cantallops, J. C. Nieves, "How to teach responsible AI in Higher Education: challenges and opportunities", *Ethics and Information Technology,* 2024, Vol. 26, No. 1, p. 3.
[65] F. Roumate, "Ethics of Artificial Intelligence, Higher Education, and Scientific Research", [in] *Artificial Intelligence in Higher Education and Scientific Research: Future Development* (Ed. Fatima Roumate), Springer Nature, Singapore 2023.
[66] D. Şenocak, A. Bozkurt, S. Koçdar, "Exploring the Ethical Principles for the Implementation of Artificial Intelligence in Education: Towards a Future Agenda", [in]

*Transforming Education With Generative AI: Prompt Engineering and Synthetic Content Creation*, IGI Global, Hershey (USA) 2024, pp. 200–213.

[67] L. Weidener, M. Fischer, "Artificial Intelligence in Medicine: Cross-Sectional Study Among Medical Students on Application, Education, and Ethical Aspects", *JMIR Medical Education,* 2024, Vol. 10, No. 1, pp. 1–18 of Article #e51247.

[68] R. T. Williams, "The ethical implications of using generative chatbots in higher education", *Frontiers in Education,* 2024, Vol. 8, pp. 1–8.

[69] I. Nuryana, B. Sugeng, E. Soesilowati, E. S. Andayani, "Critical thinking in higher education: a bibliometric analysis", *Journal of Applied Research in Higher Education,* February 15, 2024, Vol. 16, No. 5, pp. 2216–2231, *https://www.emerald.com/insight/content/doi/10.1108/JARHE-08-2023-0377/full/html* [2025-12-05].

[70] J. Pagán-Castaño, M. Arnal-Pastor, E. Pagán-Castaño, M. Guijarro-García, "Bibliometric analysis of the literature on critical thinking: an increasingly important competence for higher education students", *Economic Research,* 2023, Vol. 36, No. 2, pp. 1–23 of Article #2125888.

[71] A. S. Morris, R. Langari, *Measurement and Instrumentation: Theory and Application*, Academic Press, London (UK) & San Diego (USA) 2021 (3rd Edition).

[72] L. Finkelstein, "Education and training of engineers and scientists in measurement and instrumentation", *Measurement,* 1983, Vol. 1, No. 1, pp. 7–13.

[73] L. Mari, M. Wilson, A. Maul, *Measurement Across the Sciences: Developing a Shared Concept System for Measurement*, Springer Nature, Cham (Switzerland) 2021.

[74] "Understanding", [in] *Dictionary.com,https://www.dictionary.com/browse/understanding* [2025-12-05].

[75] E. C. Gordon, "Understanding in Epistemology", [in] *Internet Encyclopedia of Philosophy* (Eds. J. Fieser, B. Dowden), *https://iep.utm.edu/understa/* [2025-12-05].

[76] S. Grimm, "Understanding", [in] *The Stanford Encyclopedia of Philosophy* (Ed. E.N. Zalta) Winter 2024 Edition, *https://plato.stanford.edu/archives/win2024/entries/understanding/* [2025-12-05].

[77] *cf.* L. Finkelstein, "Widely, strongly and weakly defined measurement", *Measurement,* 2003, Vol. 34, pp. 39–48.

[78] J. Pearl, D. MacKenzie, *The Book of Why: The New Science of Cause and Effect*, Basic Books, New York 2018, Chapter 10.

[79] R. Z. Morawski, *Technoscientific Research: Methodological and Ethical Aspects*, Walter de Gruyter, Berlin & Boston 2024 (2nd Edition).

[80] R. Z. Morawski, "An application-oriented mathematical meta-model of measurement", *Measurement,* 2013, Vol. 46, No. 9, pp. 3753–3765.

[81] R. Z. Morawski, "Measurement as Abduction", *Perspectives on Science,* 2021, Vol. 29, No. 6, pp. 742–756.

[82] R. Z. Morawski, "Application-oriented meta-model of measurement uncertainty", *Measurement,* February 15, 2024, Vol. 225, pp. 1–8 of Article #114044.

[83] R. Z. Morawski, "Measurement as Abduction", *Perspectives on Science,* 2021, Vol. 29, No. 6, pp. 742–756.

[84] E. J. Larson, *The Myth of Artificial Intelligence: Why Computers Can't Think the Way We Do*, Harvard University Press, Cambridge (USA) 2021.

[85] M. Saarela, V. Podgorelec, "Recent Applications of Explainable AI (XAI): A Systematic Literature Review", *Applied Sciences,* 2024, Vol. 14, No. 19, pp. 1–111 of Article #8884, *https://www.mdpi.com/2076-3417/14/19/8884* [2025-12-05].

[86] R. Z. Morawski, "Measurement as Abduction", *Perspectives on Science,* 2021, Vol. 29, No. 6, pp. 742–756.

[87] E. Goh, *et al.* [16 names], "Large language model influence on diagnostic reasoning: a randomized clinical trial", *JAMA Network Open,* 2024, Vol. 7, No. 10, pp. 1–12 of Article #e2440969, *https://pubmed.ncbi.nlm.nih.gov/39466245/* [2025-12-05].
[88] P. Thagard, *Bots and beasts: What makes machines, animals, and people smart?*, MIT Press, Cambridge (USA) & London (UK) 2021, Chapter 8.
[89] *Asilomar AI Principles*, Future of Life Institute, 2017, *https://futureoflife.org/open-letter/ai-principles/* [2025-12-30].
[90] P. Thagard, *Bots and beasts: What makes machines, animals, and people smart?*, 2021, Chapter 8.
[91] T. L. Beauchamp, J. F. Childress, *Principles of Biomedical Ethics*, Oxford University Press, New York 1978.
[92] C. P. Snow, *The Two Cultures and the Scientific Revolution*, Cambridge University Press, Cambridge (UK) 1959.
[93] "Humanities and Science Integration", [in] *ScienceFather* October 2023, *https://popularscientist.com/humanities-and-science-integration/* [2025-12-05].
[94] A. Einstein, *Out of my later years: The scientist, philosopher, and man portrayed through his own words*, Philosophical Library, New York 1950, p. 9.
[95] D. J. Skorton, A. Bear (Eds.), *The Integration of the Humanities and Arts with Sciences, Engineering, and Medicine in Higher Education: Branches from the Same Tree*, The National Academies Press, Washington (DC) 2018, p. xi.
[96] *IDEAS: Integrated Degree in Engineering, Arts and Sciences*, Lehigh University, Bethlehem (PA, USA), *https://catalog.lehigh.edu/coursesprogramsandcurricula/interdisciplinaryundergraduatestudy/ideas/* [2025-12-05].
[97] *Human-centered Artificial Intelligence Master's Programme*, University of Gothenburg, Gothenburg (Sweden), *https://www.gu.se/en/study-gothenburg/human-centered-artificial-intelligence-masters-programme-t2hai* [2025-12-05].
[98] *Human AI: decoding human behaviour*, Vrije Universiteit Amsterdam, Amsterdam (The Netherland), *https://vu.nl/en/education/summerschool/human-ai/curriculum* [2025-12-05].
[99] *Human-Centered Artificial Intelligence*, Università degli studi di Milano, *https://www.unimi.it/en/education/master-programme/human-centered-artificial-intelligence* [2025-12-05].
[100] *Human-Centred AI Master's Programme*, *https://humancentered-ai.eu/* [2025-12-05].
[101] D. J. Skorton, A. Bear (Eds.), *The Integration of the Humanities and Arts with Sciences, Engineering, and Medicine in Higher Education: Branches from the Same Tree*, 2018, pp. 40–44.
[102] *ibid.*, p. 37
[103] *Future of Jobs Report 2023*, World Economic Forum, Geneva, May 2023, *https://www3.weforum.org/docs/WEF_Future_of_Jobs_2023.pdf* [2025-12-05], p. 39.
[104] R. Z. Morawski, *Technoscientific Research: Methodological and Ethical Aspects*, 2024 (2nd Edition).
[105] Quan-Hoang Vuong, Viet-Phuong La, Minh-Hoang Nguyen, Ruining Jin, Tam-Tri Le, "Are we at the start of the artificial intelligence era in academic publishing?", *Science Editing,* 2023, Vol. 10, No. 2, pp. 1–7.
[106] Hanchen Wang, *et al.* [29 names], "Scientific discovery in the age of artificial intelligence", *Nature,* 2023, Vol. 620, No. 7972, pp. 47–60.
[107] J. Gottweis, *et al.* [34 names], "Towards an AI co-scientist", 2025, *arXiv:2502.18864,* [2025-12-05].